# Personality Detection of Applicants And Employees Using K-mode Algorithm And Ocean Model

(https://arxiv.org/submit/4652175)

Binisha Mohan

*Department of Computing and Informatics*

*Bournemouth University*

Bournemouth, United Kingdom

s5531898@bournemouth.ac.uk

Dinju Vattavayalil Joseph

*Department of Computing and Informatics*

*Bournemouth University*

Bournemouth, United Kingdom

s5432223@bournemouth.ac.uk

Bharat Plavelil Subhash

*Department of Computing and Informatics*

*Bournemouth University*

Bournemouth, United Kingdom

s5530193@bournemouth.ac.uk

*Abstract*— The combination of conduct, emotion, motivation, and thinking is referred to as personality. To shortlist candidates more effectively, many organizations rely on personality predictions. The firm can hire or pick the best candidate for the desired job description by grouping applicants based on the necessary personality preferences. A model is created to identify applicants' personality types so that employers may find qualified candidates by examining a person's facial expression, speech intonation, and resume. Additionally, the paper emphasises detecting the changes in employee behaviour. Employee attitudes and behaviour towards each set of questions are being examined and analysed. Here, the K-Modes clustering method is used to predict employee well-being, including job pressure, the working environment, and relationships with peers, utilizing the OCEAN Model and the CNN algorithm in the AVI-AI administrative system. Findings imply that AVIs can be used for efficient candidate screening with an AI decision agent. The study of the specific field is beyond the current explorations and needed to be expanded with deeper models and new configurations that can patch extremely complex operations.

## I. Introduction

The distinctive patterns of thoughts, feelings, and behaviours that set one individual apart from others are referred to as personality. Understanding personality is crucial in a variety of fields, including sentiment analysis, personalised recommendations, and hiring procedures [1]. Finding a person's talent, limitations, temperament, and leadership style is the aim of the personality detection approach. Organizations can use personality identification systems to narrow down their pool of candidates, and feedback on their performance enables prospective candidates to improve their weak points. Career changers' professional objectives may be refocused with the use of this self-knowledge.

The HR department of an organisation invites an applicant for an interview based on their resume in the conventional approach of hiring. To determine whether an applicant is qualified for the position, the HR staff manually evaluates the skills listed on their resume. They conduct interviews, and the panel is tasked with determining which applicant is most qualified for the position. The manual assessment of each candidate is challenging and responding to questions is similar to taking a survey, which frequently removes the candidate's authenticity and restricts her/him to a paper document. Additionally, all hired candidates must have the proper attitude and discipline for the job, they not only assess a candidate's talents during the interview but also their personality.

Here, a model is put out to identify applicants' personalities so that organizations can find qualified individuals through a person's facial emotion, speech emotion, and restart analysis and to detect employers' personalities so that employees' well-being can be guaranteed.

As a result, a system is proposed that analyzes the personality of a person using the K-mode algorithm and Ocean Model[2]. The suggested work uses video analysis from an interview to categorize a person's performance. The Ocean model or the big five models based on AVI (referred to as AVI-AI) are suggested methods in this article for predicting a person's personality[3]. AVI-AI methods have drawn a lot of interest in the disciplines of computer sciences and human resources, particularly for autonomously evaluating personality traits [4] and communication skills [5]. The selection process is more impartial and fair than using human recruiters, which favours better and more varied candidates.

Automated interviewing reduces the administrative burden on hiring teams, enhancing flexibility, efficiency, and automation of administrative tasks. By removing pointless processes, they assist businesses in finding top personnel more quickly. The following are some advantages of automated interviews:

1. Time management

By skipping through candidates who are unqualified for the position, an automated interview helps recruiters avoid wasting time on ineffective prospects.

2. Lowers hiring expense

By enabling recruiting teams to collaborate remotely and assess candidate responses, automated interviews lower the cost of hiring.

3. Offers adaptability

Automated interviews give both candidates and recruiters more freedom since they let them customize answers at a time and place that works for them.

4. Effectively manages a high volume

By opting to analyze some passive job candidates later—a tiresome procedure otherwise—recruiters can manage large numbers of them.

5. Increases consistency of evaluation

Automated interviewing enhances evaluation consistency by establishing an organized and objective hiring procedure, assisting recruiters in quickly identifying the greatest fit.

6. Reliability & Versatility

The entire economy has relied heavily on remote work to remain afloat during the pandemic. Startups that fully embraced the ethos of working from home as a result, and the flexibility automated video interviews offer are ideal for such businesses. The interview screening videos can continue where they left off during the employment process, even if businesses choose for a hybrid work culture or fully reopen premises.

7. Efficiency

Automated video interviews are efficient in more ways than one. They not only make it possible to schedule several interviews at once quickly, but they can also do it anywhere. As a result, the business can utilize the skills of a worker who is employed elsewhere in the world but is unable to relocate for a variety of reasons.

The personality prediction is supported in this work by a questionnaire-based investigation. Openness to criticism, flexibility, team spirit, aspirations, and work ethics are among the traits that personality interview questions reveal. This aids to figure out how well a candidate may collaborate and work with team members. The responses to these queries give insight into the qualifications for the position The K-Modes clustering method is used in this survey-based investigation. The technique, which is simple to use and effective with vast amounts of data, is used to group categorical data. Based on the number of comparable categories between data points, clusters are defined. The k-modes clustering algorithm is an advancement over the k-means clustering method. K-means is the most widely used centre-based partitional clustering technique. Huang extends the k-means clustering method to the k-modes clustering algorithm to organize the categorical data:

(i) Categorize objects using a simple matching dissimilarity measure

(ii) Substituting modes for cluster means, and

(iii) Updating the modes using a frequency-based technique.[6]

To assess how well current employees are working, individual work performance (IWP), a useful and regularly used outcome measure, is often utilized. Job performance may be correlated with personality. The phrase "behaviours or acts that are related to the aims of the organization" [7] is a definition of IWP. IWP thus emphasizes employee behaviours or activities rather than the outcomes of those behaviours. Additionally, behaviours should be in the individual's control, omitting those that are limited by the environment [8]. The personalities of the employees at any given time can be ascertained from their answers to a series of questions that can be given to them. The first dimension, task performance, traditionally has received the most attention and can be defined as "the proficiency with which individuals perform the core substantive or technical tasks central to his or her job" [7]. The second dimension of IWP is contextual performance, defined as "behaviours that support the organizational, social and psychological environment in which the technical core must function" [9]. The third dimension of IWP is counterproductive to work behaviour, defined as "behaviour that harms the well-being of the organization" [8].

The study's final finding is obtained by combining the expected results from the two methodologies. The suggested system assists in selecting the most qualified candidates, improving employee quality and retention.

## II. RELATED WORK

Research efforts discussed in this article discuss various aspects of personality prediction using text, image, and video analytics. [10-12].

According to the research by [13], this model of personality can be used to evaluate multiple job performance criteria—job aptitude, learning competency, and employee information across different occupational groups, including experts, officers, supervisors, sales associates, and skilled people. This study focused on the relationships between the five personal qualities and their relationship with job performance assessments for various occupations. Using a recognized taxonomy, it investigated the connection between personality and expected levels of job performance. Additionally, it showed that there were various connections among personality traits, professions, and performance standards according to that taxonomy. A model was proposed by [14] that aid HR departments in certain organizations employ or select the best aspirant for the favoured job function. The prototype method, known as the Applicant Personality Prediction System Utilising MachineLearning, make personality predictions about candidates based on information from their resumes and from personality tests. By rating the resume on specific parameters, a personality prediction system chooses the best candidate for the specified job description. The primary goal of APPS was to determine if a candidate is suited for a job profile or not. The logistic regression algorithm provides the personality type of a candidate and the proportion of resumes that match job descriptions. The system compared the candidate's CV and job description, showing the percentage of matches so that it may be determined whether or not to hire the candidate based on that proportion.

Relying on the personality test that is provided during the registration process, APPS evaluated the characters of the applicants. Multinomial logistic regression was utilized to assess the candidate's personality, and natural language processing was used to compare the job description and résumé. APPS were constructed utilizing the logistic regression technique and decreased the hiring department's effort and time commitment.

To estimate scores, several artificial neural networks (ANNs) were trained on a sizable labelled dataset. It utilised a cascade of ANNs to forecast personality traits from static face images to examine the connections between signals from stationary facial expression cues and self-reported personality traits. The finding provides strong evidence that multidimensional personality profiles can be predicted using ANNs trained on substantial labelled datasets from static facial photos. According to the study, advanced computer vision algorithms can be used to realise personality traits in real-world photos obtained in unexpected situations. shows unequivocally that each of the Big Five features is connected to a collection of face clues that can be gathered by using machine learning methods. [15]

Nivetha et al. [16] used the Conventional Neural Network (CNN) Algorithm as well as the face spacing proportion for the online questionnaire. During online interviews, primarily, the CNN algorithm was utilised to determine a person's personality based on their face photos [17–18]. The above approach makes use of an input image dataset from a Blog (https://www.kaggle.com). Instead of preserving the actual image, the dimensions of images are saved as pixels. The above aspect of strategy for the prediction of a person's personality is based on the ratio of their face width to height. Furthermore, the set of data for the survey-based assessment from Kaggle.com [19] was employed in this research. To determine a person's personality, each situational question in the dataset received a response from 1015342 different people.

The Heterogeneous Information Ensemble (HIE) framework was employed [20] to approximate consumers' personality characteristics using diverse information, such as identity, likeness, symbols, and replying attitudes. It addressed the issue of whether the use of one's self-language is the only valuable factor of a person's personality. Based on how words were employed in different semantic categories, a psychological lexicon known as LIWC has been utilised to determine a user's personality. In addition to cutting-edge lexicon-based methods like LIWC, writers also employed To analyse the written text, Text-CNN, Correlation analysis, and pouch clustering were used. The standard CNN structure has a restricted ability to read the message because there is no clear link between the adjoining dimension in ingrained word vectors. The authors have applied the pooling layer and max pooling layers to this problem using a visual model [21]. The survey [22] demonstrates how automation and open-vocabulary reaches to natural language recognition were employed to predict a person's character based on how they talked during an interview process. The following are the results of data collected from over 46,000 people who participated in wide interview sessions and employed a HEXACO-based personality model. Following lexical studies, Ashton and Lee [23, 24] developed the six-factor Construct personality model.

The six different traits developed were openness, agreeableness, extraversion, emotionality, and conscientiousness. [25] proposes a system for automating the task of separating candidates in a recruitment process based on eligibility requirements and personality evaluation. To fulfil these requirements, an internet application was created that empowers candidates to gain entry to their details and then have Quizzes conducted to evaluate their personalities. The computer then likens the Curriculum Vitae trained sets of data that have been uploaded to assess the professional eligibility. This system employed the "Logistic Regression" machine learning technique, which helped it make good hiring choices. The ultimate personality test results will consequently be discussed with both applicants and the owner. The k-means clustering machine learning technique [26] is also used to classify personalities. The Jungian Type Inventory was developed to categorize people into personality types. As a preliminary finding, the testing revealed that the k-means model had an inertia value of 107. The data are grouped into 16 clusters. The three primary characteristics of "k-means" that make it productive and successful for the work at hand are frequently turned into significant drawbacks in the grand scheme of things. Dispersion is used exactly as a measure of scattering, while Euclidean [27] is used exactly as a metric.

The number of clusters k is regarded as an input parameter, and choosing k incorrectly can lead to undesirable outcomes. Making a diagnostic examination to ascertain the number of clusters is crucial for this reason. A paradox could result from convergence to a local minimum.

Evaluation of the best approaches for automated personality detection, which include advanced machine learning algorithms with a focus on multi-modal methods, a variety of data processing datasets, and potential uses [28]. The paper also explored that the most specific attributes for unimodal personality detection come from the visual modality. Combining inputs from multiple modalities frequently increases prediction accuracy. The accuracy was found to significantly increase when common sense knowledge and psycho-linguistic features were combined. This investigation only considered computational methods and excludes psychological studies on personality detection because it encompasses such a broad and varied topic as personality detection.

The study [29] evaluated by comparing job candidates' assessment of fairness in the AVI configuration versus the AVI setting and human ratings and response behaviours between the SVI and AVI settings using an AI judgement agent (AVI-AI). In comparison like most video interviewing investigations over the decades, that has embraced a survey questionnaire perspective [30], this study went further and combined dynamic capabilities theory and social interaction theory to explore the effects of various recorded interview features on employment selection. To fulfil the research need for clarifying how technological advances affect user reactions, hence using use a novel experimental design to examine how both candidates' and raters' reactions are impacted by asynchronous and synchronous video interviews [31]. The synchronisation of the recorded interview settings was found to be a new mediator of the connection between the first impression and discussion score. In this survey, behaviour exhibited interview (AVI) computation and a Machine learning AI engine were used to start creating an end-to-end AI conducting interviews system. The above system performs automatic personality recognition based on features extracted from the AVIs and the genuine personality scores from the respondents' self-reported survey questions and facial gestures. Employers can later evaluate sound records using this method [32]. Based on the above studies to determine a person's personality, we have employed K-Model clustering and the OCEAN model to predict the personalities.

## III. PROPOSED SYSTEM

In recent years, video interviewing has provided many advantages to recruitment businesses, including the capacity to interview candidates from anywhere in the world. Face-to-face interviews are an appropriate testing method for systematically evaluating interpersonal communication abilities [33]. While video interviews are an effective technique for locating candidates from around the world, they also present some difficulties for hiring managers and recruiters. One such challenge is interpreting body language in a virtual job interview.

This article makes suggestions for using the Ocean model based on AVI (referred to as AVI-AI) or the Big Five personality traits to forecast a person's personality [3]. AVI-AI methods have drawn a lot of interest in the disciplines of computer sciences and human resources, particularly for autonomously evaluating personality traits [4] and communication skills [5]. Unknown are the reliability and accuracy of the ground-breaking employment selection tool known as AVI-AI. Automatically conducting interviews at a specific time is possible with the help of asynchronous video interviewing technology (AVI). The interview can be reviewed by employers at a later time. If employers want to examine how the candidates replied to the questions, they may also allow anyone to watch the recorded interview. It is difficult for human reviewers to accurately.

The asynchronous video interview (AVI) was made to allow job seekers to log in, record their answers to predetermined interview questions using a webcam and microphone on a computer or mobile device, and afterwards have those responses assessed by human raters[35]. Anywhere, at any time, candidates can use AVI to record their answers to questions. Additionally, the selection process can be sped up by the human raters independently sharing and evaluating the interview video records without scheduling an interview[36].

A questionnaire-based study is also used in this work to support personality prediction. The K-Modes clustering algorithm is used for this questionnaire-based analysis. Combining the outcomes anticipated by the two approaches yields the study's ultimate conclusion.
The proposed system helps in finding the best applicants which boost employee quality and retention. You can more effectively evaluate a candidate's ability and personality via personality testing. It also helps to examine a candidate's likelihood of remaining in the position and fitting into the company's culture.

## IV. OCEAN MODEL

One of the most widely used models to describe The OCEAN Model, commonly referred to as the Big Five Personality Traits, is one of the most popular models for describing and evaluating a person's personality. Offices and other work settings are where the OCEAN model is most frequently applied. The majority of study focuses on using the big five personality traits to forecast a person's productivity and interpersonal skills at work. Understanding the five elements of the OCEAN model more thoroughly can help managers and staff build more trusting workplaces and improve interpersonal interactions [5].

The usual classification of a person's main five personality qualities is Openness, Conscientiousness, Extraversion, Agreeableness, and Neuroticism [37].

A. *Openness*
The quality best encapsulates how open-mindedness is often understood. Intellectually curious, imaginative, and open-minded traits are often found in people.

B. *Conscientiousness*
Conscientiousness has the personality trait of caution. Being conscientious implies a desire to work hard to achieve a task and a seriousness about other people's responsibilities. Conscious individuals prefer to be efficient and structured.

C. *Extraversion*
One can assess someone's extraversion to determine how active, sociable, and friendly they are. Typically social beings, extraverts focus their energy on other people and consequently the rest of the world because they get their energy from being around others.

D. *Agreeableness*
Being agreeable is having the ability to put other people's demands ahead of your own.

E. *Neuroticism*
People who exhibit neuroticism often experience negative side effects, including irritation, melancholy, self-consciousness, anxiety, and wrath. They go through such things as a result of their unstable emotions.

Once the company has determined the requirements of the position, it can use the AVI-AI-based system which uses OCEAN Model to evaluate candidates' various personality qualities. To support its analysis, a questionnaire-based study using the K-modes clustering algorithm is also used.

## V. K-MODE CLUSTERING

KModes clustering is one of the unsupervised Machine Learning algorithms that is used to cluster categorical variables. It is easy to implement and efficiently handles a large amount of data. A Kmodes technique uses a randomly selected starting cluster centre (modes) as a seed, which results in issue clustering results that are frequently dependent on the initial cluster centre of choice and the possibility of obtaining non-repeatable cluster structures. To replace "means" through "modes" in the clustering of categorical data, the K-Modes technique has been extensively used. To cluster categorical data, Huang developed a straightforward matching method in the K-Modes technique [38].

The two categories of clustering techniques are hierarchical and partitioned clustering. The two additional categories of hierarchical methods are bottom-up and top-down, and the two categories of partitioning clusters are k-mean and k-modes. By substituting the k-modes algorithm for the k-means algorithm for categorical data, the k-modes algorithm introduces a new dissimilarity metric and updates the modes using a frequency-based approach [39,40,41]. Finding homogenous groupings of objects in a given dataset is the basic definition of the clustering problem.

The following are the steps for k-Modes-based clustering (Huang 2008):[42]

1. Select the k initial mode

2. Allocate the observation to the closest cluster based on a simple dissimilarity measure. Update each cluster

mode after each allocation

3. After all the observations have been allocated to a cluster, check the dissimilarity value of each observation against the mode. If an observation turns out that the closest mode is in another cluster, move the observation to the appropriate cluster and update the mode of both clusters

4. Repeat step 3 until none of the observed changes to other clusters

How to choose the optimal number of clusters

To determine the optimal number of clusters, the Elbow method is used but it is modified to use within cluster difference. From the results of plotting within cluster differences for various values, the principle of the Elbow method takes the value of k at the point when the value does not decrease significantly with the addition of the value of k.

$$k\ m\ wc = \Sigma\Sigma\alpha v\ j=1\ i=1$$

Where WCD is the within-cluster difference, k is the number of clusters, m is the number of observations in each cluster, c is the centroid of the cluster, and d is the simple dissimilarity measure.

Mathematics Formula

Suppose that $X=\{X_1,X_2,…,X_n\}$ is a set of n object and $X_i=\{X_{i1},X_{i2},…,X_{im}\}^T$ where m

denotes the variables and I denote i

the cluster.

The Measure of Similarity

The general formula for the measure of similarity is denoted as follows.

$$d(X_i,Z_l)=\sum_{j=1}^m \delta(x_{ij},z_{lj})$$

Where $Z_l=\{z_{l1},z_{l2},…,z_{lm}\}^T$ is a prototype for cluster l. A measure of similarity for numerical variables is well-known as the euclidian distance is denoted as follows.

$$d(X_i,Z_l)= \sqrt{(\sum_{j=1}^{m_r} (x_{ij}^r-z_{lj}^r)^2)}$$

Where $x_{ij}^r$ is a value of numerical variables $j, z_{lj}^r$ is the average of prototype for numerical variables j cluster m, and the number of numerical variables.

While a measure of similarity for categorical variables is denoted as follows.

$$d(X_i,Z_l)= \gamma_l \sum_{j=1+1}^{m_c} \delta(x_{ij}^c,z_{lj}^c)$$

Where simple matching similarity measure for categorical variables is denoted as follows.

$$\delta(x_{ij}^c,z_{lj}^c)=\{(0, \& x_{ij}^c=z_{lj}^c @ 1, \& x_{ij}^c \neq z_{lj}^c)\dashv$$

Where $\gamma_l$ denotes the weight for categorical variables for cluster l that is the standard deviation of numerical variables in each cluster. The $x_{lj}^c$ denotes the categorical variables, $z_{lj}^c$ is the mode for variables j cluster l, and $m_c$ denotes the number of categorical variables.

The modification of simple matching similarity measure is as follows.

$$\delta(x_{ij}^c,z_{lj}^c)=\{(1-\omega(x_{ij}^c,l), \& x_{ij}^c=z_{lj}^c @ 1, \& x_{ij}^c \neq z_{lj}^c)\dashv$$

The above formula increases the object similarity within the cluster with categorical variables so that the result will be better where $\omega(x_{ij}^c,l)$ denotes the weight for $x_{ij}^c$ where

$$\omega(x_{ij}^c,l)=f(x_{ij}^c \mid c_l)/(|c_l| \cdot f(x_{ij}^c \mid D))\ [43]$$

Categorical data cannot be clustered using the K-means clustering method due to the different metrics it uses. The K-mode cluster algorithms are based on the K-mean pattern but eliminate the restriction on numerical data while maintaining their efficacy. By removing the restriction imposed by Kmeans after modification, this K-mode technique extends the K-mean pattern to cluster categorical data:

For categorical data objects, use simple match dissimilar evaluation or hamming distance. Alter the means of the cluster by modes.

The following processes make up the K-modes, which presuppose that the information regarding the number of probable groups of data (i.e. K) is accessible: -

1. Select K initial cluster centres, one for each cluster, and generate K clusters by randomly picking data objects.

2. Designate the data object's cluster of origin.

A technique for computing initial modes for the K-mode clustering algorithm to cluster categorical data sets is presented in the publication [44].

There are several techniques to gather evidence in the unsupervised learning context: (a) combining the outputs of various clustering algorithms; (b) producing different outputs by resampling the data, as in bootstrapping techniques (like bagging) and boosting; and (c) repeatedly executing an algorithm with various initializations or parameters.

## VI. DATASET DESCRIPTION

This work makes use of a dataset from Kaggle (https://www.kaggle.com/). An automated video interview has pre-set questions that you can respond to by recording. It analyses the facial expressions which are stored in the form of image pixels and the vocal spectrum which is stored in the form of sound signals while answering the pre-set questions.

This work makes use of the dataset for the questionnaire-based analysis that was downloaded from the Kaggle website [10]. This dataset includes responses from 1015341 individuals for each situational question intended to predict a person's personality. 50 distinct question types are included in the dataset as columns, including "I am quiet around strangers," "I feel at ease around others," "I don't like to bring attention to myself," and more [35].

The responses to the situational questions in this dataset are used to determine a person's personality. The scenario-based interview questions are shown in Fig.1. The database with the recorded responses to the scenario-based questions is depicted in Fig. 2.

```
1. What motivates you?
   ⊙1  ⊙2  ⊙3  ⊙4  ⊙5

2. Tell me about a stressful situation and how you handled it.
   ⊙1  ⊙2  ⊙3  ⊙4  ⊙5

3. Is your preference to work on a team or on your own? Why?
   ⊙1  ⊙2  ⊙3  ⊙4  ⊙5

4. If you could change one thing about yourself, what would it be and why?
   ⊙1  ⊙2  ⊙3  ⊙4  ⊙5

5. If you could be any animal, what would you be?
   ⊙1  ⊙2  ⊙3  ⊙4  ⊙5
```

Fig1. Scenario-based Questions

| Name | Applicant ID | Scenario 1 | Scenario 2 | Scenario 3 |
|---|---|---|---|---|
| DIYA B | AID111 | 2 | 2 | 2 |
| ANUSHREE D | AID102 | 3 | 3 | 3 |
| DHARSHIKA V | AID105 | 4 | 3 | 4 |
| MONISHA S | AID107 | 2 | 4 | 3 |
| SNEHA K | AID177 | 2 | 2 | 2 |
| PRIYA M | AID134 | 3 | 3 | 3 |
| DIVYA | AID101 | 4 | 5 | 1 |
| YAMUNA C | AID131 | 3 | 3 | 5 |
| BANUPRIYA C | AID128 | 3 | 3 | 5 |

Fig2. Recorded data for Scenario-based Questions

The personality of the employees in an organization can also be evaluated based on the Kmode clustering algorithm. For this, the Questionnaire on Individual Work Performance was analyzed.

A certain questionnaire is given to the employees based on OCEAN Model in addition to questions that help to evaluate their performances. The data are grouped into clusters to predict the personality of the workers in an organization. Fig.3. shows an example of questions asked to find out the personality of employees and their attitude towards their work.

```
1. Tell me about a time you had to fill in for someone. Were you successful? How did the experience make you feel?
   ⊙1  ⊙2  ⊙3  ⊙4  ⊙5

2. Do you prefer working in a team or on your own? Why?`
   ⊙1  ⊙2  ⊙3  ⊙4  ⊙5

3. Tell me about a time your manager wasn't satisfied with the results of your work. How did you discuss the issues and what did you do differently the next time?
   ⊙1  ⊙2  ⊙3  ⊙4  ⊙5

4. If you could change one thing about yourself, what would it be and why?
   ⊙1  ⊙2  ⊙3  ⊙4  ⊙5

5. What are you passionate about?
   ⊙1  ⊙2  ⊙3  ⊙4  ⊙5
```

Fig.3. Individual Work Performance-based Questions.

| Name | Employee ID | Scenario 1 | Scenario 2 | Scenario 3 |
|---|---|---|---|---|
| ANU | EID1 | 3 | 2 | 2 |
| MOHAN | EID5 | 2 | 3 | 3 |
| SRIYA | EID2 | 4 | 1 | 2 |
| JAISON | EID7 | 2 | 2 | 3 |
| RIYA | EID8 | 2 | 3 | 2 |
| SEEMA | EID3 | 3 | 3 | 3 |

Fig.4. Recorded data for Individual Work Performance-based Questions.

The responses from Individual work performance based are used to predict employees' personalities. It can also be used to determine whether the employee is still fit to work in the organization. His/her attitude towards the company changes or not with time can also be analysed. How the employees need to be trained to bring out their overall output and how to make the employees work in a stress-free atmosphere can also be known by the organization.

Personality plays a significant role in organisational behaviour because it has a large impact on how people think, feel, and act at work. Individuals' personalities have an impact on their group behaviour, attitudes, and decision-making processes. Interpersonal skills have a big impact on how people act and react in workplace situations. In the workplace, personality affects motivation, leadership, performance, and conflict. Managers are better prepared to succeed and accomplish their goals the more they are aware of the part that personality plays in organisational behaviour. [47].

## VII. DATA ANALYSIS

The data was examined using the dataset for image pixels. The dataset is prepared theoretically, but it needs to be normalized before it can be used with Python tools.

Sklearn is used to preprocess the data from the questionnaire dataset [7]. The dataset is created from the raw data after the data have undergone analysis. The unlabeled dataset is fed into the K-Modes algorithm, which clusters categorical data.

## VIII. PREDICTING PERSONALITIES

The AVI is a clever administrative system, but there is no evidence to date that it is intelligent or has analytical capabilities. But analysis of this data is necessary [29]. AI-based AI needs to be trained properly. In a study by Hickman et al. in 2021, an AI was instructed to analyse a list of items to automatically assess a person's personality [37].

| Behaviour | Definition |
|---|---|
| Verbal - What interviewees say; the content of their response | • Word count |
| | • Proportion of words longer than six letters |
| | • 70 Linguistic Inquiry and Word Count (LIWC) dictionaries |
| | • n-grams with n = 1, 2 (i.e., words and two-word phrases) |
| Paraverbal - How interviewees sound when delivering their responses | • Pitch |
| | • Jitter |
| | • Frequency |
| | • Shimmer |
| | • Loudness |
| | • Harmonics-to-Noise ratio |
| | • Alpha ratio |
| | • Hammarberg index |
| | • Spectral slope |
| | • Loudness peaks per second |
| | • Length of continuously voiced regions |
| | • Length of continuously unvoiced regions |
| | • Voiced segments per second |
| Nonverbal - What interviewees do (e.g., facial expressions, posture) | • Head pose |
| | • 19 facial action units activation intensity |
| | • Mean |
| | • Standard deviation |
| | • Kurtosis |
| | • Skewness |
| | • Facial action unit cooccurrences |

Fig3. List of things that an AI was asked to process in a study by Hickman et al in 2021

The result obtained from this kind of AI may not be accurate. The accuracy depends on the data we provide to the AI. The stored data from AVI-AI can use the Conventional Neural Network (CNN) Algorithm to predict the personality from video streams. The analysis of image pixels in video datasets can be obtained with the use of the CNN Algorithm. The frequency, pitch, loudness, and various vocal aspects were evaluated for each answer against the stored data in AI. AVI predicts the personality based on the Big Five personality traits or the OCEAN Model.

Later this data is combined to predict the personality with the help of results obtained from the questionnaire-based analysis. Here also, the clusters of data were formed based on Big Five personality traits or OCEAN Model to find out the personalities. The goal of the unsupervised learning technique known as clustering is to group the population of data points into several groups where each group's data points are more similar to one another and distinct from those in the other groups. The gathering of objects is essentially dependent on how similar and unlike they are to one another [45].

One unsupervised machine learning approach used to cluster categorical variables is Kmodes clustering. The most commonly used unsupervised machine learning is Kmeans but here we use Kmodes clustering instead of Kmeans.

Continuous data are clustered using KMeans using distance as a mathematical metric. The more closely related our data points the closer the distance. This means changes in centroids.

The distance cannot be calculated for categorical data points, though [46]. KModes algorithm is what we choose. It makes use of the differences between the data points (total mismatches). Our data points are more comparable overall, the smaller the differences. Rather than using means, it employs modes [36].

This algorithm groups the responses to scenario-based questions into clusters based on the big five personality traits. The results of the two approaches outlined above can be combined to determine a person's highest personality.

## IX. RESULTS AND DISCUSSION

(A) Applicant Analysis

With the use of the CNN algorithm in the AVI-AI administrative system, a person's personality was predicted based on OCEAN Model. The accuracy obtained was 92.6% based on our analysis.

It varies with training parameters. Fig. shows a pie chart representing personality percentages based on OCEAN Model or Big Five personality traits from the above analysis.

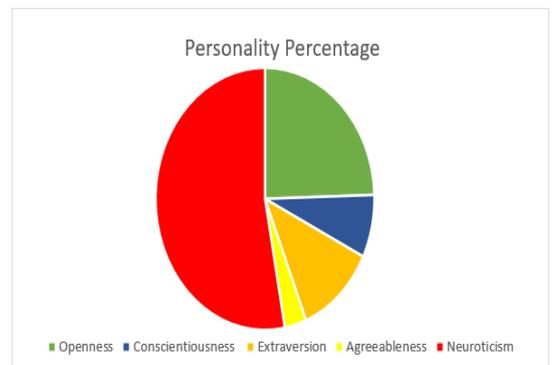

Fig.6. Pie chart showing personality percentage using CNN-based AVI-AI administrative system

The K-Modes clustering algorithm produced output with 91.2% accuracy based on responses from the questionnaire-based analysis.
The personality percentage produced using the K-Modes clustering approach is shown in Fig. 7 as a pie chart.

K-Modes clustering is more accurate than using K-means clustering as the K-mode algorithm uses categorical data to form clusters.

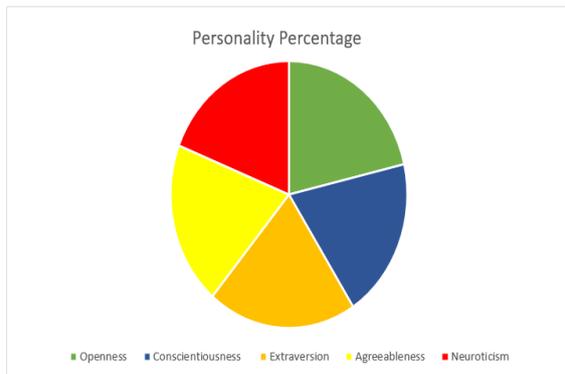

Fig.7. Pie chart showing personality percentage from questionnaire-based analysis

The big five personality traits of openness, conscientiousness, extraversion, agreeableness, and neuroticism are used to predict a person's behaviour based on the result obtained from AVI-AI and scenario-based questions. Fig. 8 shows the personality percentage produced by combining both methods.

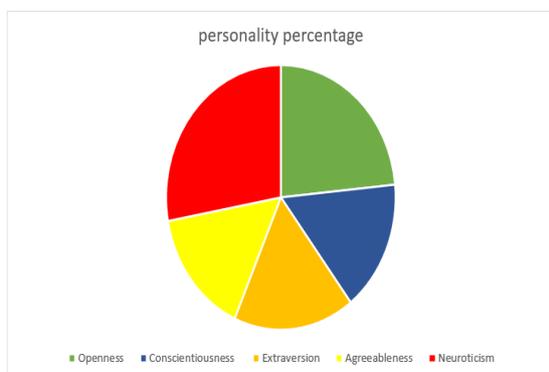

Fig.8. Pie chart showing personality percentage

Thus, the combination of personality predicted using the CNN algorithm[16] in AVI-AI based on the OCEAN Model and the answer of the subject to scenario-based or situational questions led to the discovery that the neuroticism personality percentage was higher than that of other personalities.

Employers can gauge an employee's ability to fit within the organization through personality testing. Less time and effort would be required to fill the position since they would be more likely to stay with the organization for a very long period if they are a better fit.

Employers or recruiters use personality tests as a sort of assessment to help them filter applicants and locate people with personality attributes that ideally match their businesses. It is intended to reveal specific facets of candidates' personalities and predict whether or not they will succeed in particular roles [47].

Role-specific exams are frequently expected to be a component of the interview process for specialized positions in recruitment, such as computer engineers or data analysts. For candidates at the mid-level and above, however, personality tests have emerged as a new component of the process since they give employers more precise insights into the traits, abilities, and behaviours of the applicants [48].

B) Employee Analysis

With the use of the Kermodes clustering algorithm, the responses towards individual work performance-based questions
are analysed to predict the personality of employees and their attitude towards work is determined. As a result, the employer will be able to learn more about the employee's attitudes toward their jobs and the company culture. Various unique views have an impact on their personalities. A person reacts to a situation depending on their values and characteristic traits. These features are established throughout a particular timeframe and cannot be easily changed, supervisors should attempt to comprehend this and then prevent it. [48]

An individual with openness, emotional control, and agreeableness is likely to have fewer disputes, collaborate effectively in teams, and have a good mindset toward their profession. Positions that require teamwork or leadership should be accessible to those with this personality type. Those without these traits will be less motivated and more negative under comparable situations.

Positive interpersonal skills are a personality quality that significantly impacts the workplace. People with this feature typically love working with others, and they possess the empathy and sensitivity necessary to get along with others. People who possess this quality frequently work in positions that need them to interact with clients, oversee staff, or resolve conflicts [47].

Independence and decision-making are significantly influenced by personality. While characteristics like neuroticism and narrow-mindedness make it difficult to make wise decisions alone and under pressure, others like ego, conscientiousness, and pro-activeness do. Managers can place individuals with these characteristics in settings where they can perform at their highest level.
Individuals with specific traits are much more encouraged when they are given tasks that better match them. Their overall achievement at work is impacted as they are satisfied every day. The above influences workplace productivity because a motivated workforce and favourable attitude lead to greater accomplishments[47].

To make hiring decisions and create elite teams, organisations are increasingly utilising cutting-edge tools like personality testing. One of the numerous advantages of personality tests is that they help employers find the proper candidates by letting them know about a candidate's motivations, values, and work preferences for a particular job role. The benefits of taking personality tests in the hiring process are that the same information can be utilised throughout an employee's career to develop them for a suitable role in addition to being a wonderful tool to screen candidates at the screening stage [49].

There is no denying that personality tests have advantages for companies. They give insightful information about how an applicant's personality will affect their behaviour at work, enabling recruiters to comprehend how a prospect will collaborate with others, solve issues, and control their emotions at work [50].

Several advantages of personality tests include [50]:

Faster hiring procedure:
   It speeds up and simplifies the hiring process. The number of unwanted interviews can be avoided by choosing the best prospects from the pool of applicants during the online evaluation process. The recruitment team occupies less time because personality test results are available instantly.

Improved Understanding of Candidates:
   Employers can identify the necessary personality traits in potential candidates by using objective evaluations. The employer can choose a group of candidates based on underlying personality traits for a certain job position.

Dark Personality Traits to Look Out For:
   Employers might alter their hiring practices by utilizing the results of the personality test to find undesirable personality traits such as opportunism, self-obsession, lack of sensitivity, temperament, and impulsive behaviour.

Removes Bias:
   The use of personality assessments greatly reduces the possibility of unconscious bias when choosing applicants for a job position. By assessing candidates according to how well they perform in the relevant interpersonal skills, employers may make fair hiring decisions. A candidate with the necessary skills would be the best for the job. Numerous factors are taken into consideration while evaluating applicants for employment, and the personality assessment produces clear results.

Cost-Effective:
   Personality tests are inexpensive and easy to use. A reliable personality test can provide data about an individual that would take a company months to find out in a couple of hours. These tests provide a positive investment return by lowering or lessening your chances of making bad recruits. In addition to being a costly investment, poor recruitment can significantly harm the productivity and morale of your other employees.

Several disadvantages of personality tests include [50]:

Adaptation for a specific task:

   The same personal characteristics are not necessary for every job description. To be valid and dependable, a personality test must be adaptable enough to be tailored to different employment requirements and responsibilities.

For instance, an individual with a convincing character would be a great fit for a sales position. However, a position in digital marketing does not necessitate that skill. Personality assessments must be personalised to consider the skills and qualities that each employer needed from its employees.

Although they performed flawlessly on the personality test, this does not necessarily imply that they are the best candidate for the position or that they will perform with the same effect if they are given the job. This emphasises once more how crucial it is to not merely hire people based solely on their personality traits.

Various Personality Tools on the Market:
   There is a wide range of character diagnostics available tools that assume to provide accurate results. However, these assertions may be a little overstated. Incorrect outcomes from an imperfect personality test may result in a candidate personality type that would make it challenging for the company to select the candidate who is the best for the job.

High-quality content

   Unsure questionnaires lead to the poor content of several personality tests. The finished reliability of the results is affected by that. It's indeed essential to confirm and get to understand the professionals who will be evolving the content for your personality assessments.

   Personality assessments require dependable personality testing, which also calls for competent information. Companies must select the best aspirants and provide them with the necessary training to provide reliable personality assessments.

Response Formats:

   Attendees in psychological testing commonly react in a manner that's also socially acceptable rather than demonstrating their truthful personalities. Those who can modify their comments in response to produce the desired outcomes for the business, try to deceive hiring managers into making illogical decisions. Institutions should choose a personality test that considers and account for information for these considerations.

Potential Reaction:

Numerous potential employees prefer not to start taking the short quiz out of fear, a lack of real-world experience, or too many time constraints. Evaluations quite often overlook crucial elements like ethnic heritage or cultural differences, which leads to the loss of possibility.

The above list includes the both benefits and drawbacks of using an objective psychological test. Institutions and HR staff should make sensible decisions when selecting personality assessments because they'll use them to make repeated crucial related to recruitment and development-related judgements.

Before actually deciding regardless of whether to use a psychological test in the application process or not, a corporate must weigh the benefits and drawbacks in the company's best interests. Nevertheless, as discussed in the cons, it may also deter anybody from applying for a job in certain sectors and certain roles in special. As has been previously mentioned, this same test should not be utilized as the only metric for hiring since doing so could be detrimental to the company in the long run.

## X. CONCLUSION AND FUTURE SCOPE

By analyzing a person's facial expression, speech inflexion, and resume, a model is developed to identify applicants' personality types so that organizations can find suitable individuals. A similar model is created to determine employers' personality types so that the well-being of their employees can be guaranteed. In the current generation, AVIs and AI algorithms are becoming more and more common and are used extensively in many industrial and occupational sectors to find the right candidate. There is, however, little evidence that AVI and AI decision agents can improve employment selection quickly and affordably. This study uses the big five personality traits to examine the effects of AVI on interview ratings and job applicants' behaviours. Here individual's personality is predicted using the CNN algorithm in the AVI-AI administrative system using the OCEAN Model and the K-Modes clustering algorithm is implemented for predicting employee well-being including work pressure, working environment and relationship with peers. Using the ocean model, our analysis' accuracy for choosing the best applicant was 92.6% and its accuracy for questionnaire responses was 91.2%. Based on the results from AVI-AI and scenario-based questions, the big five personality traits of openness, conscientiousness, extraversion, agreeableness, and neuroticism are used to predict a person's behaviour. combination of personality predicted using the CNN algorithm in AVI-AI based on the OCEAN Model and answer of the subject to scenario-based or situational questions led to the discovery that the neuroticism personality percentage was higher than that of other personalities. These findings imply that AVIs can be used for efficient candidate screening with an AI decision agent. To better understand whether AI assessment results can impact human decision-making in employment selection, researchers are comparing the reliability and validity of AI decision agents and human raters. The study of the specific field expands with deeper models and new configurations that can patch extremely complex operations. Shortly, we intend to create more sophisticated systems that utilize AVI-AI techniques to analyze conversations between users and data from online media to diagnose and treat users' mental processes.